# Entropy production in ac-calorimetry

J.-L. Garden* and J. Richard

*Institut Néel, CNRS et Université Joseph Fourier, BP 166, 38042 Grenoble Cedex 9, France.*

**Abstract**

*In calorimetry and particularly in heat capacity measurements, different characteristic relaxation time constants may perturb the experiment which cannot be considered at thermodynamic equilibrium. In this case, thermodynamics of irreversible processes has to be taken into account and the calorimetric measurements must be considered as dynamic. In a temperature modulated experiment, such as ac-calorimetry, these non-equilibrium experiments give rise to the notion of frequency dependent complex heat capacity. In this paper, it is shown that for each irreversible process an experimental frequency dependent complex heat capacity can be inferred. Furthermore, we demonstrate rigorously that a same equality connects the imaginary part of these different complex heat capacities with the entropy produced during these irreversible processes. Finally, we claim that the presence of an imaginary part in the measured heat capacity always indicates that a certain amount of heat does not participate to the classical equilibrium heat capacity of the sample when measured over the observation time scale.*

* Corresponding author. Fax: 33 (0) 4 76 87 50 60

*E-mail address:* jean-luc.garden@grenoble.cnrs.fr (J.-L. Garden)

# 1. Introduction

It is well-known that calorimetric experiments can be perturbed by different parasitic relaxation time constants. These different time constants can alter the measurement in such a way that what we measure is not what we really believe. For example, in heat capacity measurements, one of the fundamental time constant is $\tau_{ext}$, the time constant of the adiabaticity. This is the external relaxation time constant of the temperature of the sample towards the constant temperature of the bath. If the time scale of the measurement is larger than this time constant, the calorimetric measurement cannot be considered as adiabatic (in a calorimetric sense and not in a thermodynamic sense) and heat has time to relax towards the thermal bath. A correction has thus to be taken into account considering the heat exchange coefficient in order to correctly derive the heat capacity of the sample. The second important relaxation time constant is still due to the non-equilibrium behaviour of the temperature of the whole sample. It is connected to the diffusion of heat within the body of the sample. What is thus the exact temperature of the whole sample when the thermal diffusivity is low? Suppose that in a modulated temperature calorimetric experiment the frequency of the input power is so high that at the other extremity of the sample the thermometer never oscillates. We can understand that, in this case, the exact heat capacity of the sample is never recorded. The third time constant that we want to consider in this article is the kinetic relaxation time constant of specific internal degrees of freedom of the sample. When heat is supplied to the sample in a fast way, some of these degrees of freedom have never time to absorb this quantity of heat over the time scale of the experiment. In this case, these degrees of freedom do not contribute to the heat capacity measured by the experimentalist. In modulated temperature measurements this particularity has provided the famous notion of frequency dependent complex heat capacity with a real and an imaginary part satisfying the Kramers-Kronig dispersion relations (see for example, the following reviews and references therein [1-3]). This latter notion has been already investigated in the literature of calorimetry and we do not want to discuss this in details here. Nevertheless, we will recall that the imaginary part of the complex heat capacity is, in this case, also deeply connected to the entropy produced over one period of the temperature cycle. The last of these relaxation time constants that we want to address is not widely known and is linked to the relaxation of the thermal power due to finite velocity of the heat carriers. It involves a regime where the Fourier's law becomes inexact.

In this paper, we demonstrate that for each of this characteristic time constants, there is a different irreversible process where a different complex heat capacity can be inferred. In these four different cases, the imaginary part of these complex heat capacities is always connected to the entropy produced over one period of the temperature cycle during the irreversible process. The paper is composed of the following different sections involving each time one of the time constants aforementioned. Before going into the details of these different sections, we would like firstly, to explain what we want to say by "time scale of the measurement", which is another very important time constant in experimental calorimetry.

## 2. Time scale of the measurement

The time scale of the measurement is the smallest characteristic time interval during which a physical parameter of a system is recorded by the experimentalist without any specific averaging. Over this time interval, an experimental point can be inferred. In classical calorimetric experiment, for example in differential scanning calorimetry (DSC), this time interval is the smallest finite time interval $\Delta t$ during which an experimental heat capacity point (more precisely a differential heat flow point) is recorded. The measured heat capacity is thus the natural averaging of the instantaneous heat capacity taken over this time interval. In modulated temperature measurements, this characteristic time is the period of the oscillating input thermal power. The influence on the measurement of the other time constants depends on the ratio of their own value as compared to this characteristic time scale. The time scale of the measurement is the reference against which the various time constants encountered in the calorimetric experiments have to be compared. Let us take a well-known example: when we consider the kinetic relaxation time constant due to slow structural change inside a sample or the slow advancement of a chemical reaction, the ratio of this time constant on the time scale of the measurement is called the Deborah number [4]:

$$D = \tau/\Delta t \tag{1}$$

This typical ratio is used to characterize the difference between a liquid and a glassy state. For infinitely fast time scale of the measurement all is frozen ($D \rightarrow +\infty$) and nothing has time to move, we observe a frozen-in solid. On the contrary, under an observation time scale which tends to the infinity all is in movement and we observe a liquid ($D \rightarrow 0$). In calorimetric modulated temperature experiments, the Deborah number $\omega\tau$, appears in the denominator of the frequency dependent complex heat capacity.

## 3. External thermal relaxation time constant of the temperature

*3.1. Definition.*

Let us consider the figure 1, where a classical finite thermodynamic system with a heat capacity $C$ is linked via a heat exchange coefficient $K$ to a thermal bath with a constant temperature $T_0$. The macroscopic thermodynamic system is a sample under calorimetric investigation. Its temperature is well defined and in this section we consider that its thermal diffusivity is infinite. The external thermal relaxation time constant of the temperature of the system defines the temperature of equilibrium as compared to that of the heat bath. It represents the time constant necessary for the heat to relax towards the heat sink. At thermodynamic equilibrium the temperature of the system equals precisely those of the bath. On the other hand, the temperature of the system can be constant and different from the temperature of the bath when stationary conditions are fulfilled. The system is then in a constant non-equilibrium state often called a stationary steady-state. The ratio of the external thermal time constant on the time scale of the measurement defines the condition of adiabaticity of the measurement. According to the value of this ratio, the calorimetric experiment may be realized in an adiabatic manner or not. The calorimetric experiment is adiabatic (in a calorimetric sense and not in a thermodynamic sense) if there is not heat exchanged between the sample and the heat bath, other than the quantity of heat supplied to (or released from) the sample by the experimentalist during the time scale of the measurement. If the experiment is not adiabatic, heat has time to flow away from the sample during this characteristic time. Subsequently, during this time scale the temperature of the sample relaxes exponentially. This adiabaticity time constant is defined by:

$$\tau_{ext} = C/K \qquad (2)$$

$C$ is the heat capacity of the sample and $K$ the coefficient of heat exchange.

*3.2. Principle of the ac-calorimetry method.*

In this section we shall briefly recall the principle of the ac-calorimetry method which will serve as a model for our demonstration, although all the developments made in this paper can be applied with more or less complications to all other dynamic calorimetric methods.

An input thermal power $P(t) = P_{dc} + P_{ac}$ constituted by a dc and an ac term is supplied to the system. In the stationary regime, the temperature response is composed by a dc and an ac component:

$$\begin{cases} \Delta T_{dc} = T_{dc} - T_0 = \dfrac{P_0}{K} \\ T_{ac}^* = \delta T_{ac} \exp[i(\omega t - \varphi)] \end{cases} \quad (3)$$

The star indicates complex notations of the oscillating variables used here for the sake of calculus simplicity (for example the exact oscillating temperature of the system is the real part of $T_{ac}^*$: $T_{ac} = \text{Re}(T_{ac}^*) = \delta T_{ac} \cos(\omega t - \varphi)$). $T_{dc}$ is the mean constant dc temperature of the sample, $T_0$ is the constant temperature of the bath, $\delta T_{ac}$ is the amplitude of the oscillating temperature, $\omega$ is the angular frequency and $\varphi$ is the phase between the oscillating temperature and the input oscillating heat flow with a phase taken by convention equal to zero:

$$P_{ac}^* = P_0 \exp(i\omega t) \quad (4)$$

When the period $2\pi/\omega$ is the only characteristic time involved in the measurement, the heat capacity is simply:

$$C_{mes} = \frac{P_{ac}^*}{dT_{ac}^*/dt} = \frac{P_{ac}^*}{i\omega T_{ac}^*} = \frac{P_0}{\omega \delta T_{ac}} \tag{5}$$

with a phase lag of $\pi/2$ between the thermal power and the temperature.

*3.3. Complex heat capacity*

When the heat exchange coefficient cannot be neglected in the measurement (no adiabatic measurement) the temperature of the sample obeys in this case to the following differential equation:

$$P(t) = C\frac{dT}{dt} + K(T - T_0) \tag{6}$$

Considering only the oscillating part of this equation in the stationary regime (the dc part is given in the equation (3)) the equation can be transformed in:

$$\tau_{ext} \dot{T}_{ac}^* + T_{ac}^* = \frac{P_{ac}^*}{K} \tag{7}$$

In the stationary regime, the resolution of this equation gives directly the oscillating temperature:

$$T_{ac}^* = \frac{P_{ac}^*}{K(1+i\omega\tau_{ext})} = \frac{P_{ac}^*}{K+i\omega C} \qquad (8)$$

We observe the appearance of the adiabaticity ratio, $\omega\tau_{ext}$, on the denominator which is a direct indication of the strength of the calorimetric adiabaticity of the measurement. This equation simply means that the oscillating temperature is the sum of two perpendicular components. An experimental complex heat capacity can be derived from the definition (5):

$$C_{mes} = \frac{P_{ac}^*}{i\omega T_{ac}^*} = \frac{K+i\omega C}{i\omega} = C - i\frac{K}{\omega} = C' - iC'' \qquad (9)$$

Therefore, considering only the adiabaticity time constant, the measured heat capacity is a complex number. The real part of the complex heat capacity is the heat capacity of the sample. The imaginary part has the dimension of a heat capacity. In fact it is a thermal conductance by unit of angular frequency. It is linked to the heat lost over the time scale of the measurement. As we have mentioned before, its value depends directly on the ratio of the thermal relaxation time on the time scale of the measurement, $\omega\tau_{ext}$. This is however an irreversible thermodynamics process because heat flows out of the sample irreversibly.

*3.4. Entropy production*

When the adiabaticity time constant plays a role, the sample can not be regarded as a thermally insulated thermodynamic system. Hence, the heat bath (or thermal bath) has to be taken into consideration in the balance of the entropy produced during this thermodynamic non-equilibrium process (see figure 2). We are in presence of a single thermodynamic system composed by two homogeneous discrete sub-systems. One is the sample and the other the heat bath. A thermodynamic sub-system is homogeneous if there is no gradient of intensive parameters wherein. Exchanges of extensive parameters between each sub-system are simply due to differences of intensive parameters such as the pressure (volume exchange), the chemical potential (matter exchange) and evidently the temperature (heat exchange). For homogeneous discrete sub-systems the calculus of the entropy produced in the entire system due to exchange of extensive parameters between each sub-part are thus very simple. In the present case, the entropy produced in the entire system (sample + bath) is only due to the exchange of heat between the sample and the heat bath (see figure 3).

In the following, we assume that the stationary conditions are fulfilled. That is to say, the dc temperature has reached a constant value $T_{dc}$ (or this value varies so slowly that its rate can be neglected). This value corresponds to the dc temperature of the stationary steady non-equilibrium state. Since the temperature of the system is the sum of a dc and an ac component, the entropy production can be separated in two contributions. The dc part is due to the mean constant heat flux exchanged between the sample and the bath. There is a dc temperature gradient between the sample and the bath (see fig. 4). It is the reason why a stationary non-equilibrium steady-state is reached at the level of the sample. The ac term is linked to the heat loss towards the bath due only to the oscillatory part of the temperature. Let us see how these two terms can appear and can be separated. Let us envisage the case of a thermal power supplied to the sample (see fig. 4). Thus, the temperature of the sample is greater than those of the heat bath, $T = T_{dc} + T_{ac} > T_0$. Hence, since it has enough time, heat relaxes irreversibly from the sample to the heat sink.

From the figure 5, the amount of heat involved in the ac-calorimetry experiment can be separated in two different types. At the level of the sample, there is external heat exchanged between the sample and the surroundings ($dQ_e^S$) due to the heat flow supplied by the experimentalist, and an internal heat exchange due to the relaxation towards the bath ($dQ_i^S$). At the level of the bath, there is only an internal heat exchange flowing from the sample ($dQ_i^B$). Obviously, we have the two following relations:

$$dQ^S = dQ_e^S + dQ_i^S \tag{10}$$

which is just the expression of the conservation of energy at the level of the sample, and where $dQ_e^S$ is positive if heat is supplied to the sample from the outside world, and $dQ_i^S$ is negative because heat flows from the hot to the cold points. We have also:

$$dQ_i^S + dQ_i^B = 0 \tag{11}$$

which simply means that what is released from the sample is taken by the bath. Afterwards, considering the entire system "sample-bath", the total entropy variation is written:

$$dS_{tot} = \frac{dQ^S}{T} + \frac{dQ^B}{T_0} = \frac{dQ_e^S}{T} + \frac{dQ_i^S}{T} + \frac{dQ_i^B}{T_0} = \frac{dQ_e^S}{T} + dQ_i^S \left( \frac{1}{T} - \frac{1}{T_0} \right) \tag{12}$$

This expression can be separated in two contributions. One is external and must be positive or negative depending on either heat is supplied to the sample or released from the sample by the experimentalist. The other contribution is definitely positive and called internal entropy variation inside the system. It is only this contribution which is connected to the irreversible process due to the heat flow from the sample towards the heat bath.

Let us now envision this latter term in details:

$$d_iS = dQ_i^S \left( \frac{1}{T_{ac}+T_{dc}} - \frac{1}{T_0} \right) \approx -dQ_i^S \frac{(\Delta T_{dc}+T_{ac})}{T_0^2} \tag{13}$$

The two temperature differences $\Delta T_{dc}$ and $T_{ac}$ have been neglected as compared to $T_0$ (recall that in ac-calorimetry $T = T_{dc} + T_{ac} = T_0 + \Delta T_{dc} + T_{ac}$). Subsequently, the instantaneous rate of production of entropy is:

$$\sigma_i = \frac{d_iS}{dt} = -\frac{dQ_i^S}{dt}\frac{(\Delta T_{dc}+T_{ac})}{T_0^2} \tag{14}$$

In ac-calorimetry from Sullivan and Seidel work [5], it is well-known that the heat flux exchanged between the sample and the heat bath via the heat exchange coefficient $K$ is the sum of two components, a dc and an ac term included in the second term of the right-hand side of (6):

$$-\frac{dQ_i^S}{dt} = K\Delta T_{dc} + KT_{ac} \tag{15}$$

Consequently the entropy production takes the following expression:

$$\sigma_i = \frac{K}{T_0^2}(\Delta T_{dc}+T_{ac})^2 \tag{16}$$

which can be separated in two components. The first is a dc component:

$$\sigma_i^{dc} = K\left(\frac{\Delta T_{dc}}{T_0}\right)^2 = P_0 \frac{\Delta T_{dc}}{T_0^2} = \frac{P_0^2}{KT_0^2} \tag{17}$$

This term represents the constant and continuous entropy produced inside the system due to the dc constant heat flow between the sample and the heat bath in the stationary regime. This heat flow is exactly compensated by the dc thermal power supplied by the experimentalist to the sample, maintaining it in a non-equilibrium stationary state. The second term is the sum of two oscillatory terms (one oscillates at the frequency of the input power and the second at twice the frequency):

$$\sigma_i^{ac} = K\left(\frac{T_{ac}}{T_0}\right)^2 + 2K\frac{\Delta T_{dc}T_{ac}}{T_0^2} \tag{18}$$

This term is the instantaneous entropy production due to the oscillatory component of the sample temperature relaxing towards the bath. Now, if we take the average of this entropy production over one period of the temperature cycle, then it remains only the contribution of the twice frequency oscillating term:

$$\bar{\sigma}_i^{ac} = K\left(\frac{\delta T_{ac}}{T_0}\right)^2 \int_{-T/2}^{T/2} \cos^2(\omega t - \varphi)dt = \pi\left(\frac{\delta T_{ac}}{T_0}\right)^2 \frac{K}{\omega} \tag{19}$$

With the equation (9) of the complex heat capacity we have:

$$\bar{\sigma}_i^{ac} = \pi\left(\frac{\delta T_{ac}}{T_0}\right)^2 C'' \tag{20}$$

Hence, in one period of the temperature modulation there is positive creation of entropy due to oscillatory heat exchange between the sample and the heat bath, which is proportional to the imaginary part of the experimental frequency dependent complex heat capacity. To be more precise, knowing that the modulus of the oscillating temperature can be expressed as:

$$\delta T_{ac} = \frac{P_0}{\omega |C_{mes}|} \tag{21}$$

and also that the amount of heat involved per half-period of the oscillating cycle is:

$$\delta Q_0 = \int_{-T/4}^{T/4} P_0 \cos(\omega t) dt = 2\frac{P_0}{\omega} \tag{22}$$

then over one period of the temperature oscillation (20) can be expressed as follows:

$$\bar{\sigma}_i^{ac} = \frac{\pi}{4} \frac{\delta Q_0^2}{T_0^2} \frac{C''}{|C_{mes}|^2} = \frac{\pi}{4} \frac{\delta Q_0^2}{T_0^2} \operatorname{Im}\left(\frac{1}{C_{mes}}\right) \tag{23}$$

Hence, the entropy produced irreversibly per period of the temperature modulation due to oscillatory exchange of heat between the sample and the thermal bath is directly proportional to the imaginary part of the complex impedance of the measurement. During this period of time, we can say that heat is lost (dissipated, absorbed) because it does not contribute to the measurement of the usual heat capacity of the sample. All this last formula have been already derived by different authors who start in deriving the entropy at thermodynamic equilibrium to the second order term in the oscillatory temperature [2, 6, 7]. Nevertheless, as it was clarified in a recent paper [3], this derivation has nothing to do with the well-known classical expression of the frequency dependent complex heat capacity where internal degrees of freedom are involved (see the last section). Here, this approach concerns the ac-calorimetry case, but the TMDSC method will be envisaged under the same point of view in a forthcoming publication. Indeed, in TMDSC method the condition of adiabaticity is basically not fulfilled, because generally heat is directly supplied to the sample from the heat bath via the heat exchange coefficient $K$.

**4. Internal thermal relaxation time constant of the temperature**

*4.1. Definition.*

As in the previous section, this thermal time constant is also related to the thermal disequilibrium of the sample. In this case, the thermodynamic system that we have to consider is only composed by the sample which is thermally insulated from the heat bath (the condition of adiabaticity is supposed to be respected). The ratio of this thermal time constant on the time scale of the measurement defines the condition of homogeneity of the temperature of the sample. That is to say, according to the value of this ratio, the temperature may be or may not be the same anywhere and at any time within the sample. The calorimetric experiment fulfils the condition of temperature homogeneity of the sample if during the time scale of the measurement heat is not diffused (or absorbed) along the spatial dimensions of the sample. If the requirement is not fulfilled, heat is lost along the path linking the hot point (generally the heater) and the cold point of the sample (usually the thermometer). In this case, as in all diffusion phenomena, the temperature measured at the level of the thermometer relaxes exponentially over a spatial dimension. Let us point out that not only the finite value of the diffusivity of the sample medium is a limiting factor, but also all the thermal interfaces (thermal contacts) encountered between the hot source and the thermometer are limiting factors for a perfect internal temperature equilibrium of the sample. Let us now enter in the general treatment of complex heat capacity measured in diffusive media.

*4.2. Semi-infinite diffusive medium*

Generally, the case of semi-infinite diffusive medium is the simplest and pedagogical example to treat diffusion of heat from the Fourier's law in oscillatory regime. Here we used this model for simplicity keeping in mind the objective that we want to reach, but the more complicated ac-calorimetry case is treated in the appendix. Let a semi-infinite homogeneous medium thermally coupled to a thermal bath of constant temperature $T_0$ (cf. figure 6). Let us suppose a heater supplying an ac thermal power $P_{ac}^*$ at the "free face" of the system at the origin of the one dimensional spatial axis ($x = 0$). In the oscillatory regime and forgetting the dc term for simplicity, the oscillatory temperature at a distance $x$ from the heater obeys to the diffusion equation:

$$\frac{\partial T(x,t)}{\partial t} = D \frac{\partial^2 T(x,t)}{\partial x^2} \tag{24}$$

where $D$ is the thermal diffusivity of the medium:

$$D = \frac{k}{\rho c} \tag{25}$$

where $k$ is the thermal conductivity, $\rho$ is the density and $c$ is the bulk specific heat. The ac stationary solution of this spatio-temporal variables equation is [8]:

$$T_{ac}^* = \delta T_0 \exp\left(-\frac{x}{\lambda}\right)\exp\left[i\left(\omega t - \frac{x}{\lambda} - \frac{\pi}{4}\right)\right] \tag{26}$$

where $\delta T_0$ is the amplitude of the oscillating temperature at the origin, and $\lambda$ is the characteristic diffusion length of the temperature within the sample:

$$\lambda = \sqrt{\frac{2D}{\omega}} \tag{27}$$

The phase $-\pi/4$ is due to the boundary condition $P = P_0 \exp(i\omega t)$ at $x = 0$. The Fourier's law establishes the relation between the heat flux propagating inside the sample and the temperature gradient at a distance $x$ from the origin:

$$P(x,t) = -k\frac{\partial T}{\partial x}(x,t) \tag{28}$$

In the oscillatory and stationary regimes this heat flux is:

$$P_{ac}^* = P_0 \exp\left(-\frac{x}{\lambda}\right)\exp\left[i\left(\omega t - \frac{x}{\lambda}\right)\right] \tag{29}$$

where $P_0$ is the amplitude of the alternative power at the origin which is linked to $\delta T_0$ by the following equation:

$$P_0 = \frac{k\sqrt{2}}{\lambda}\delta T_0 = \sqrt{k\rho c\omega}\delta T_0 \qquad (30)$$

Now, integrating the diffusion equation (24) from 0 to the infinity (semi-infinite medium) we obtain:

$$S\int_0^{+\infty}\frac{\partial P_{ac}^*}{\partial x}dx = -S\int_0^{+\infty}\rho c\frac{\partial T_{ac}^*}{\partial t}dx \qquad (31)$$

which gives:

$$-SP_0\exp(i\omega t) = -S\rho ci\omega\delta T_0\exp\left[i\left(\omega t - \frac{\pi}{4}\right)\right]\int_0^{\infty}\exp\left[-\frac{x}{\lambda}(1+i)\right]dx = -C_\lambda i\omega\delta T_0\frac{\exp\left[i\left(\omega t - \frac{\pi}{4}\right)\right]}{1+i}$$

(32)

where $C_\lambda = \rho c S\lambda$ is a characteristic heat capacity obtained on a characteristic volume given by the product of the surface $S$ of the sample and the characteristic diffusion length $\lambda$. The natural definition of an experimental complex heat capacity is in this case:

$$C_{mes} = \frac{P_{ac}^*(x=0,t)}{\partial T(x=0,t)/\partial t} \qquad (33)$$

This complex heat capacity can be deduced for example from such an experiment realized with a thermometer placed at the same location than those of the heater. It is worth noticing that this expression is valid at any position $x$ along the $x$-dimension of the sample, because the ratio of the thermal power on the temperature time derivative is independent of *x*. With (32) it yields to:

$$C_{mes} = \frac{C_\lambda}{1+i} = \frac{C_\lambda}{2}(1-i) \tag{34}$$

It has to be remarked that in the case of diffusive semi-infinite medium, the heat capacity which can be inferred from an oscillating temperature experiment with a heater placed at the top of the sample, and a thermometer located at any distance *x* from this side, is equal to half of the heat capacity calculated from a volume of the homogeneous sample represented by the surface *S* and the characteristic thermal diffusion length *λ*. This heat capacity can be measured equally from the in-phase or the out-of-phase oscillating temperature component. It is well-known that the relaxation time constant involved in these types of situations is approximately:

$$\tau_{int} \approx \frac{L^2}{D} \tag{35}$$

With (26) we have:

$$\omega \tau_{int} \approx \frac{L^2}{\lambda^2} \tag{36}$$

Consequently the heat capacity measured in this latter experiment can be expressed as follows:

$$C_\lambda \approx \frac{C}{\sqrt{\omega \tau_{int}}} \tag{37}$$

where $C$ is the heat capacity due to the entire volume of the sample.

*4.3. Entropy production*

The instantaneous entropy production by unit of volume resulting from the irreversible aspect of the propagation of heat in diffusive media is given by:

$$\sigma_i^v = \frac{k}{T_0^2}\left(\frac{\partial T}{\partial x}\right)^2 \tag{38}$$

This formula is derived again from the product of the thermodynamic force $\vec{\nabla}_x\left(\frac{1}{T}\right) \approx \frac{\vec{\nabla}_x T}{T_0^2}$ in the linear regime of validity of the Fourier's law, with the thermodynamic induced flux, the heat flux (see (28)). Multiplying by the constant surface $S$ and integrating from zero to the infinity, it gives the instantaneous rate of production of entropy in the entire volume:

$$\sigma_i = S\int_0^{+\infty} \frac{k}{T_0^2}\left(\frac{\partial T}{\partial x}\right)^2 dx = \frac{2k(\delta T_0)^2 S}{\lambda^2 T_0^2}\int_0^{+\infty}\exp\left(-\frac{2x}{\lambda}\right)\cos^2\left(\omega t - \frac{x}{\lambda}\right)dx = \frac{(\delta T_0)^2 \omega C_\lambda}{4T_0^2}\left[1 + \frac{1}{2}\cos(2\omega t) + \frac{1}{2}\sin(2\omega t)\right]$$
(39)

Taking the time average of this latter expression over one period of the temperature cycle gives:

$$\overline{\sigma}_i = \pi \frac{(\delta T_0)^2}{T_0^2} \frac{C_\lambda}{2} = \pi \frac{(\delta T_0)^2}{T_0^2} C'' \tag{40}$$

In the appendix, we show also the validity of this relation in the particular case of ac-calorimetry in diffusive regime. We can assume that this expression is also valid for any kind of diffusive experiments with any types of sample with complicated spatial geometry.

**5. Internal thermal relaxation time constant of the heat flux**

*5.1. Beyond the Fourier's law*

Some specific situations can happen in which the Fourier's law is not valid anymore. As a matter of fact, Fourier's law yields to a paradoxical infinite speed of propagation of heat in a medium. In fact, when the ratio of the absolute temperature on the mean free path of the heat carriers becomes small as compared to the temperature gradient, the Fourier's law goes out of its domain of validity [9]:

$$\frac{1}{T}\frac{\partial T}{\partial x} \gg \frac{1}{l} \Rightarrow Fourier\ is\ not\ valid \tag{41}$$

where $l$ is the mean free path of the heat carriers. This particular situation can be reached theoretically and experimentally in studies of propagation of heat in non-homogeneous diffusive media [10]. Anyway, the discussion on the domain of validity of the Fourier's law seems to be still opened. For instance, let us suppose that a modulated calorimetric experiment is realized in such a situation. A supplementary term, taking into account the relaxation time constant $\tau$ of the heat flux (relaxation of the heat carriers) has to be added to the classical Fourier's law. This yields to the Vernotte-Cattaneo equation [11, 12]:

$$\tau \frac{d\dot{Q}}{dt} + \frac{dQ}{dt} = K_s \Delta T \qquad (42)$$

It is straightforward to see that the classical Fourier's law, where $K_s$ is the internal thermal conductance of a sample, is found when this relaxation time constant becomes negligible. The equation (42) is a first order linear equation ensuring that once again the treatment is realized in the vicinity of thermodynamic equilibrium. As usual, the temperature variation $\Delta T$ in (42) is written:

$$\Delta T = T - T_0 = T_{dc} + T_{ac} - T_0 \qquad (43)$$

If we write $P = \frac{dQ}{dt}$ the heat flux, then at equilibrium $\dot{P} = 0$ and $T_{ac} = 0$ and we recover the Fourier's law:

$$P_0 = K_s(T_{dc} - T_0) \qquad (44)$$

In order to consider disequilibrium around this constant dc situation (stationary condition), the Vernotte-Cattaneo equation can more explicitly be written:

$$\tau \delta \dot{P} + \delta P = K_s T_{ac} \qquad (45)$$

where $\delta P = P(t) - P_0$ is the little departure of $P$ around its constant equilibrium value $P_0$. The resolution of (42) in complex notations and under stationary conditions yields to:

$$\delta P^* = \frac{K_s T_{ac}^*}{1 + i\omega\tau} \tag{46}$$

Let us remark that the heat flux propagating inside the sample is the sum of two oscillating components with a phase difference of π/2. A part of the thermal power is dispersed, and the other part is absorbed due to the relaxation of the heat carriers inside the sample. From this last result, a different perspective might be to consider a complex thermal conductance inside the sample.

*5.2. Complex heat capacity*

Starting with the definition (5) of the complex heat capacity we obtain:

$$C_{mes} = \frac{\delta P^*}{i\omega T_{ac}^*} = -\frac{K_s}{\omega}\frac{\omega\tau}{[1+(\omega\tau)^2]} - i\frac{K_s}{\omega}\frac{1}{[1+(\omega\tau)^2]} \tag{47}$$

Consequently, just beyond the Fourier's law the imaginary part of the frequency dependent complex heat capacity is:

$$C'' = \frac{K_s}{\omega}\frac{1}{[1+(\omega\tau)^2]} \tag{48}$$

*5.3. Entropy production*

Let us start with the same definition (17) of the rate of production of entropy as a product of a thermodynamic force by the induced thermodynamic flux:

$$\sigma_i = \frac{d_i S}{dt} = -\frac{dQ_i^S}{dt}\frac{(\Delta T_{dc} + T_{ac})}{T_0^2} \qquad (49)$$

This time, the induced thermodynamic flux is just given by (46) and we obtain (in complex notations):

$$\sigma_i^* = P^*(t)\frac{(\Delta T_{dc} + T_{ac}^*)}{T_0^2} = \left(P_0 + \frac{K_s T_{ac}^*}{1+i\omega\tau}\right)\frac{(\Delta T_{dc} + T_{ac}^*)}{T_0^2} \qquad (50)$$

The dc rate of production of entropy which maintains the system in a non-equilibrium quasi-stationary state is found again (see equation (15))

$$\sigma_i^{dc} = K_s\left(\frac{\Delta T_{dc}}{T_0}\right)^2 = P_0\frac{\Delta T_{dc}}{T_0^2} = \frac{P_0^2}{K_s T_0^2} \qquad (51)$$

This time, the permanent heat flux is flowing inside the sample from the hot source towards the cold source, and the internal thermal conductance across the sample replaces the heat leak of the non-adiabatic case. For the ac part, all the other oscillating terms are either terms modulated at the frequency $\omega$ or terms expressed as a product of two oscillating terms in quadrature, apart for one term which oscillates at twice the frequency. When the net entropy produced over the time scale of the experiment is calculated by taking the time integral over one cycle, only this latter term contributes. It is straightforward to see that this term is:

$$\overline{\sigma}_i^{ac} = \frac{\pi}{\omega} \frac{K_s}{[1+(\omega\tau)^2]} \frac{\delta T_{ac}^2}{T_0^2} = \pi \left(\frac{\delta T_{ac}}{T_0}\right)^2 C'' \tag{52}$$

## 5. Kinetic relaxation time constant of internal degrees of freedom

When a particular internal degree of freedom is suddenly perturbed by a temperature variation, it relaxes following a characteristic kinetic relaxation time constant. This characteristic time is the cause of the so-called frequency dependent complex heat capacity or generalized calorimetric susceptibility [13, 14]. This later thermodynamic complex quantity is known for a long time ago. The frequency dependent complex heat capacity appears at the beginning of the 20$^{th}$ century in the field of ultrasonic absorption on diluted gas. Then this notation was refund later in the field of chemical relaxation and after used a lot in the famous calorimetric experiments of Birge and Nagel with the so-called 3ω calorimetric method. We would just like recall here in a summary the important physical aspect of this unusual thermodynamic quantity [1-3].

Firstly, a very important hypothesis necessary to understand well this concept is to assume that the system is in thermal equilibrium. That is to say that the first two studied previous thermal relaxation time constants do not play a role here. For ac-calorimetry experiments, mathematically this requirement implies the two following inequalities:

$$\tau_{\text{int}} << \frac{1}{\omega} << \tau_{ext} \tag{53}$$

Experimentally the useful working frequency range is chosen in such a way that the system is in a stationary regime with external temperature equilibrium (adiabaticity conditions) and with internal temperature equilibrium (infinite thermal diffusivity and perfect thermal contact). Secondly, let us now observe a particular internal degree of freedom inside the sample. This internal degree of freedom generally contributes to the total heat capacity of the sample under study. That is to say, among the quantity of heat supplied to the sample by the experimentalist, this degree of freedom can absorb the necessary amount of heat which totally excites it, allowing the system to be in another equilibrium thermodynamic state (another sample configuration, another physical state, another chemical composition, another phase, etc…). However, if heat is supplied in a shorter time interval than the kinetic relaxation time constant of the degree of freedom, this degree does not contribute entirely to the equilibrium value of the measured heat capacity under the time scale of observation (because it is still relaxing). In this situation, the measured heat capacity is a non-equilibrium quantity which varies on time. The heat capacity becomes a dynamic quantity. On a strict thermodynamic point of view, the sample is out of equilibrium. As an example, the most well-known case of irreversible process is the case of chemical reactions where the internal degree of freedom is characterized by an internal parameter, or an order parameter, usually called degree of advance of the reaction or extent of the reaction. Over a given variation of the temperature of the sample in a given time interval, it is possible that the extent of the reaction can not reach its equilibrium value during this time scale because of the slow kinetic of the chemical reaction. Sometimes, the kinetic of the internal reorganization inside the sample is so slow, that the internal degree of freedom is completely frozen. The sample is thus completely frozen-in over the time scale of observation. At this level, from an original work of Prigogine and Mazur [15], we have envisaged recently that during the relaxation of the order parameter characterizing the slow internal degree of freedom, a certain amount of heat is lost (or absorbed) along a virtual axis represented by the value of this order parameter [3]. Consequently, this amount of heat does not participate to the equilibrium part of the measured heat capacity, exactly in such a same way envisaged for irreversible heat diffusive effects and irreversible relaxation of heat carriers of the previous sections. Moreover, this relaxation is accompanied by a definite positive entropy production which, when it is averaged over the time scale of the experiment (positive entropy creation), is directly connected to the imaginary part of the complex heat capacity exactly in a same manner than in the case of non-equilibrium temperature of the sample (non adiabaticity and non homogeneity of the temperature of the sample).

## 6. Conclusion

When a time constant appears in modulated temperature calorimetric experiment, it has to be compared to the characteristic time scale of the experiment in order to see whether the experiment is reversible (at thermodynamic equilibrium) or irreversible (out of thermodynamic equilibrium). When this time constant cannot be neglected as compared to the time scale of observation, the heat capacity measurement becomes dynamic and the measured heat capacity becomes a complex number. For each time constant considered in this paper, it has been demonstrated that the imaginary part of the complex heat capacity is connected following exactly the same equality to the positive entropy produced over the time scale of observation. The presence of an imaginary part in the complex heat capacity indicates that a part of the total heat supplied to the system cannot totally excite the sum of the degrees of freedom constituting the whole heat capacity of the system. In the case of the non-adiabaticity, this amount of heat flows away from the sample via the heat leak. Since the imaginary part is inversely proportional to the thermal frequency, this effect is accentuated at low frequency. On the other hand, for all the others time constant considered (thermal diffusivity, finite velocity of the heat carriers, and slow internal degree of freedom), the higher is the thermal frequency, the bigger is this quantity of heat lost within the sample for the measurement of the static equilibrium heat capacity.

From the results obtained in this paper, we would like to ask an opened question: since the same fundamental relationship is obtained either for the time constant implied in the thermal equilibrium of the sample (thermal diffusivity and calorimetric adiabaticity) or for the non-equilibrium behaviour of slow internal degrees of freedom within the sample, may this remark give rise to a generalized definition of the temperature? May the heat capacity and particularly its imaginary part give rise to a more general definition of the temperature for systems out of thermodynamic equilibrium?

This work was realized in the team of "Thermodynamique des Petits Systèmes" and the "Pôle de Capteurs Thermométriques et Calorimétrie" of the Institut Néel. The authors would like to thank O. Bourgeois and H. Guillou for stimulating discussions and many corrections of the manuscript.

Appendix

In general, in ac-calorimetry experiments, a sample of volume $V = SL$ is linked by a thermal conductance $K$ to a thermal bath of constant temperature $T_0$ (cf. figure 7). The heater is assumed to be located at the position $x = 0$ and the thermometer at the distance $x = L$ from the heater on the sample as depicted in figure 7.

The two boundary conditions necessary for the resolution of the Fourier's diffusion equation (see equation (24)) are in this case:

$$\begin{cases} \left(\dfrac{dQ}{dt}\right)_{x=0} = P_0 - KT_{x=0} \\ \left(\dfrac{dQ}{dt}\right)_{x=L} = 0 \end{cases} \qquad (1)$$

For this geometry, the stationary solution of the diffusion equation yields to:

$$T_{ac}^*(x,t) = \frac{P_0 ch(ax-\theta)\exp(i\omega t)}{Kch\theta + K_s \theta sh\theta} \qquad (2)$$

with $K_s = k\dfrac{S}{L}$ the internal thermal conductance inside the entire sample volume and the complex parameter $a$:

$$a = \alpha L(1+i) \qquad (3)$$

and the complex parameter $\theta$:

$$\theta = (1+i)\alpha \tag{4}$$

with

$$\alpha = \sqrt{\frac{\omega C}{2K_s}} \tag{5}$$

and $C = \rho S L c$ the total heat capacity of the sample.

The complex heat capacity at the position $x = 0$ can be defined as follows:

$$C_{mes} = \frac{P_0}{\partial T / \partial t)_{x=0}} = \frac{P_0}{i\omega T^*_{ac,x=0}} = \frac{P_0}{i\omega T^*_{ac,x=L} ch\theta} \tag{6}$$

where $T^*_{ac,x=L}$ is the oscillating temperature measured with the thermometer.

From this definition, the imaginary part of the complex impedance of the measurement is calculated:

$$\mathrm{Im}\left(\frac{1}{C_{mes}}\right) = \frac{\frac{\alpha K_s}{2}(sh2\alpha - \sin 2\alpha) + K(ch^2\alpha - \sin^2\alpha)}{K^2(ch^2\alpha - \sin^2\alpha) + 2\alpha^2 K_s^2(sh^2\alpha + \sin^2\alpha) + \alpha K K_s(sh2\alpha - \sin 2\alpha)} \tag{7}$$

Considering that the experiment is realized at such a frequency than the sample is thermally insulated from the heat bath (adiabaticity condition), the latter equation is simplified by putting $K = 0$:

$$\mathrm{Im}\left(\frac{1}{C_{mes}}\right) = \frac{(sh2\alpha - \sin 2\alpha)}{4\alpha K_s (sh^2\alpha + \sin^2\alpha)} \tag{8}$$

The entropy production is calculated within the entire sample by:

$$\sigma_i = \int (Flux \times Force)dV = S\int_0^L \frac{k}{T_0^2}\left(\frac{\partial T_{ac}^*}{\partial x}\right)\left(\frac{\partial T_{ac}^*}{\partial x}\right)^{cc} dx \tag{9}$$

where the modulus of the oscillating temperature and the constant dc gradient across the sample are still together neglected before the bath temperature $T_0$. The *cc* superscript on the second temperature gradient means the complex conjugation.

The temperature gradient is obtained from (2):

$$\frac{\partial T_{ac}^*}{\partial x} = \frac{aP_0 sh(ax-\theta)\exp(i\omega t)}{Kch\theta + K_s\theta sh\theta}$$

$$= \frac{\frac{\alpha}{L}(1+i)P_0\left\{sh\left[\alpha\left(\frac{x}{L}-1\right)\right]\cos\left[\alpha\left(\frac{x}{L}-1\right)\right] + ich\left[\alpha\left(\frac{x}{L}-1\right)\right]\sin\left[\alpha\left(\frac{x}{L}-1\right)\right]\right\}\exp(i\omega t)}{[Kch\alpha\cos\alpha + \alpha K_s(sh\alpha\cos\alpha - ch\alpha\sin\alpha)] + i[Ksh\alpha\sin\alpha + \alpha K_s(ch\alpha\sin\alpha + sh\alpha\cos\alpha)]} \tag{10}$$

Thus:

$$\left(\frac{\partial T_{ac}^*}{\partial x}\right)\left(\frac{\partial T_{ac}^*}{\partial x}\right)^{cc} = \frac{2\frac{\alpha^2}{L^2}P_0^2\left\{\sin^2\left[\alpha\left(\frac{x}{L}-1\right)\right] + sh^2\left[\alpha\left(\frac{x}{L}-1\right)\right]\right\}}{K^2(ch^2\alpha - \sin^2\alpha) + 2\alpha^2 K_s^2(sh^2\alpha + \sin^2\alpha) + \alpha KK_s(sh2\alpha - \sin 2\alpha)} \tag{11}$$

The entropy production is:

$$\sigma_i = \frac{P_0^2}{T_0^2} \frac{\alpha K_s}{2} \frac{(sh2\alpha - \sin 2\alpha)}{K^2(ch^2\alpha - \sin^2\alpha) + 2\alpha^2 K_s^2(sh^2\alpha + \sin^2\alpha) + \alpha K K_s(sh2\alpha - \sin 2\alpha)} \quad (12)$$

Taking the limit when *K* tends to zero (adiabaticity) and integrating over one half a period of the modulation (because of the complex conjugation) gives:

$$\bar{\sigma}_i = \frac{\pi}{\omega} \frac{P_0^2}{T_0^2} \frac{\alpha K_s}{2} \frac{(sh2\alpha - \sin 2\alpha)}{2\alpha^2 K_s^2(sh^2\alpha + \sin^2\alpha)} = \frac{\pi}{\omega} \frac{P_0^2}{T_0^2} \text{Im}\left(\frac{1}{C_{mes}}\right) = \pi \left(\frac{\delta T_{ac}}{T_0}\right)^2 C'' \quad (13)$$

Figure 1: A simple classical finite thermodynamic system (a sample under calorimetric investigation) of heat capacity *C* at a temperature *T* is linked via a thermal conductance *K* to a thermal bath of constant temperature $T_0$.

Figure 2: The total system under interest is composed by two homogeneous sub systems, the sample and the heat bath, which are thermally coupled each other by the heat exchange coefficient *K*.

Figure 3: The entropy produced inside the system is due to the exchange of heat between the two sub-systems with different temperatures.

Figure 4: The thermodynamic system is represented beside a time versus temperature diagram. A dc constant temperature gradient is maintained between the system and the heat bath. Hence, a dc heat exchange of heat is established across the thermal link. Also an ac temperature component oscillates at the level of the sample. Hence, an ac heat exchange of heat is established across the thermal link.

Figure 5: An amount of heat $dQ_e^S$ is supplied to the sample from the outside of the system by the experimentalist. Inside the system, the amount of heat which goes away from the sample via the heat exchange coefficient *K* is entirely captured by the thermal bath ($dQ_i^S = dQ_i^B$).

Figure 6: A semi-infinite homogeneous medium is directly linked to a thermal bath of constant temperature. In a stationary condition, a heater supplied an ac thermal power at the top face of the medium located at the position $x = 0$. At a distance *x* from the top face, a thermometer records the temperature. The thermal bath is located at an infinite distance from the heater.

Figure 7: Typical situation of ac calorimetry experiment. A heater supplied an oscillating thermal power at a face of the sample and a thermometer records the temperature at the other face of the sample at a distance *L* from the heater.

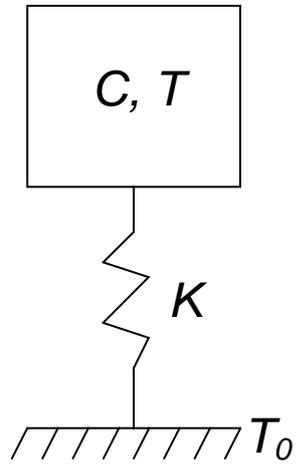

J.-L. Garden, Thermochimica Acta Fig1

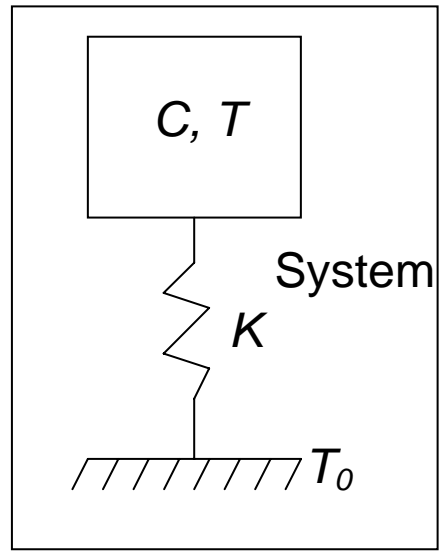

J.-L. Garden, Thermochimica Acta Fig2

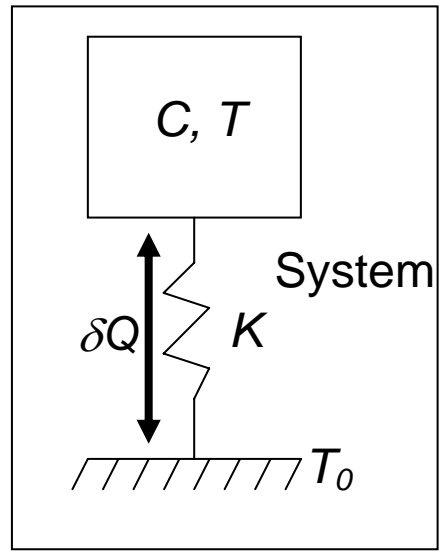

J.-L. Garden, Thermochimica Acta Fig3

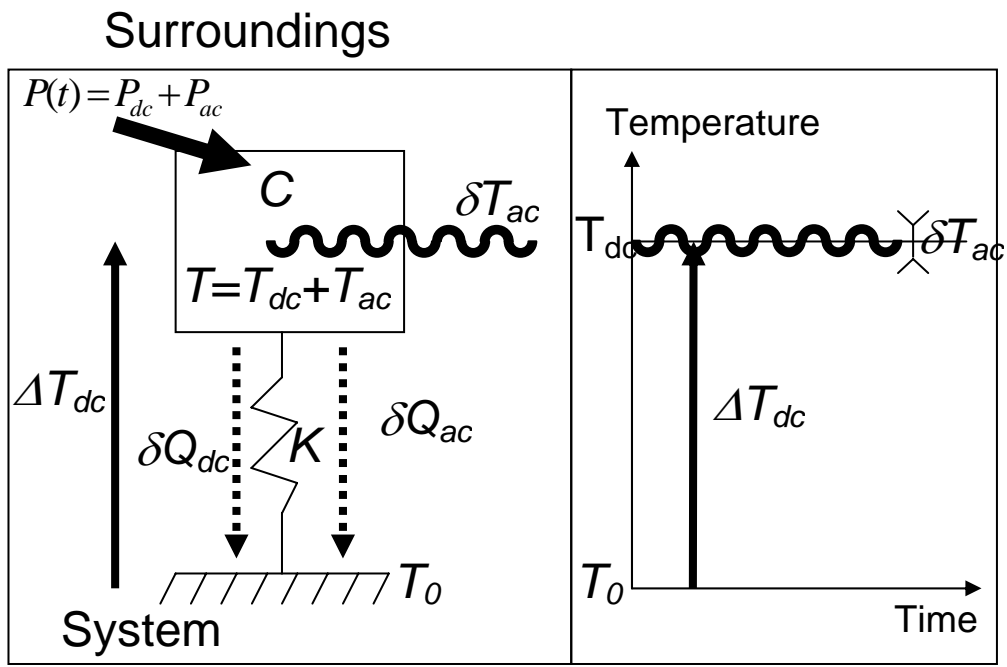

J.-L. Garden, Thermochimica Acta Fig4

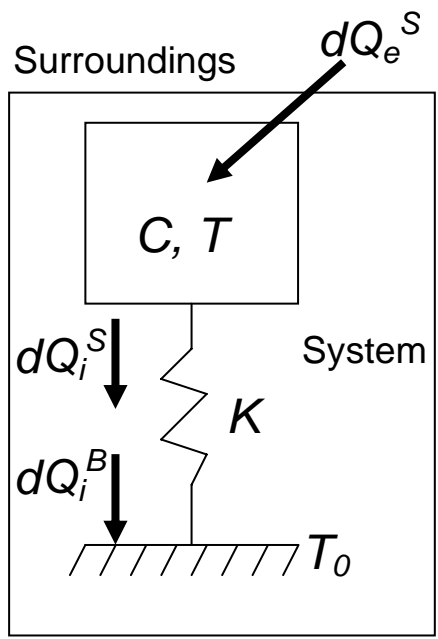

J.-L. Garden, Thermochimica Acta Fig5

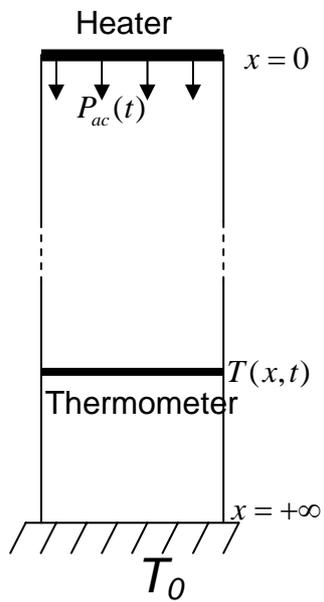

J.-L. Garden, Thermochimica Acta Fig6

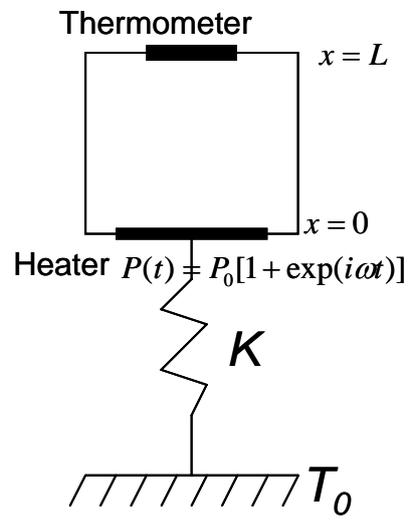

J.-L. Garden, Thermochimica Acta Fig 7